\documentclass[
    ,final            
  ]
  {aipproc}

\layoutstyle{6x9}

\newcommand {\slept}         {\tilde{\ell}}

\newcommand {\sel}           {\tilde{\rm e}}

\newcommand {\smu}           {\tilde{\mu}}
\newcommand {\stau}          {\tilde{\tau}}

\begin{document}

\title{Searches for Gauge-Mediated Supersymmetry Breaking Topologies in \boldmath{$e^+e^-$} collisions at LEP2
}


\author{Gabriele Benelli}{
  address={Department of Physics, 
	   The Ohio State University, 
	   191 West Woodruff Avenue,
	   Columbus, OH-43210-1117, U.S.A.}
}

\begin{abstract}
In gauge-mediated supersymmetry (SUSY) breaking (GMSB) models the lightest 
supersymmetric particle (LSP) is the gravitino and the phenomenology is driven 
by the nature of the next-to-lightest SUSY particle (NLSP) which is either the 
lightest neutralino, the stau or mass degenerate sleptons. Since the NLSP 
decay length is effectively unconstrained, searches for all possible lifetime 
and NLSP topologies predicted by GMSB models in $e^+e^-$ collisions are 
performed on the data sample collected by OPAL at centre-of-mass energies up 
to 209 GeV at LEP.\\
Results independent of the NLSP lifetime are presented for all relevant final 
states including direct NLSP pair-production and, for the first time, also 
NLSP production via cascade decays of heavier SUSY particles.\\
None of the searches shows evidence for SUSY particle production. Cross-section
limits are presented at the 95\% confidence level both for direct NLSP
production and for cascade decays, providing the most general, almost model 
independent results. \\
These results are then interpreted in the framework of the minimal GMSB 
(mGMSB) model, where large areas of the accessible parameter space are 
excluded. In the mGMSB model, the NLSP masses are constrained to be
$m_{\tilde\chi_1^0} > 53.5$~GeV$c^2$, $m_{\tilde\tau_1} > 87.4$~GeV$c^2$ and 
$m_{\tilde\ell} > 91.9$~GeV$c^2$ in the neutralino, stau and slepton co-NLSP
scenarios, respectively.\\ 
A complete scan on the parameters of the mGMSB model is performed, 
constraining the universal SUSY mass scale $\Lambda$ from the direct SUSY 
particle searches: $\Lambda>40,\,27,\,21,\,17,\,15$\,TeV/$c^2$ for messenger 
indices $N=1,\,2,\,3,\,4,\,5$, respectively, for all NLSP lifetimes.

\end{abstract}

\maketitle


\section{Introduction}

Supersymmetry\cite{susy}, one of the proposed solution to the hierarchy 
problems of the Standard Model (SM), postulates the existence of a bosonic 
partner for each SM fermionic particle and viceversa. The discovery of these 
superpartners would be the most direct evidence for SUSY. Since these 
particles are not observed to have the same mass as their SM partners, SUSY 
must be a broken symmetry. In the most widely investigated scenarios, it is 
assumed that SUSY is broken in some {\em hidden} sector of new particles and 
is {\em communicated} to the {\em visible} sector of SM and SUSY particles by 
gravity or gauge interactions.

We present a study of gauge-mediated SUSY breaking topologies~\cite{opal_gmsb} 
using the data collected by the OPAL detector at LEP up to the highest 
center-of-mass energies of 209 GeV.

\section{Searches for GMSB topologies}

An attractive feature of GMSB models is that the hidden sector can lie at 
masses as low as $10^4$~GeV$/c^2$. In most current GMSB theoretical work
\cite{theo}, it is assumed that this sector is coupled to a 
messenger sector, which in turn couples to the visible sector through normal 
SM gauge interactions.

The minimal GMSB model introduces five new parameters and a sign:  the SUSY
breaking scale ($\sqrt{F}$), the SUSY particle mass scale ($\Lambda$), the
messenger mass ($M$), the number of messenger sets ($N$), the ratio of the 
vacuum expectation values of the two Higgs doublets ($\tan\beta$) and
the sign of the Higgs sector mixing 
parameter (sign($\mu$)).

In GMSB models the LSP is a light gravitino ($m_{\rm \tilde G}<$~1~MeV$/c^2$), 
and the nature of the NLSP, which is either the lightest neutralino 
($\tilde\chi^0_1$), stau ($\tilde\tau_1^{\pm}$) or 
mass-degenerate sleptons ($\sel_R^{\pm}$, $\smu_R^{\pm}$ and 
$\stau _1^{\pm}$), determines the phenomenology. As the gravitino couples 
very weakly to heavier SUSY particles, these will decay typically in a 
cascade to the NLSP which then decays via either $\tilde\chi^0_1 \rightarrow 
\gamma \tilde\mathrm{G}$ or $\tilde\ell^{\pm} \rightarrow \ell^{\pm} 
\tilde\mathrm{G}$. 
We study~\cite{opal_gmsb} all relevant final states: both direct NLSP 
production and its appearance in the decay chain of heavier SUSY particles, 
like charginos, neutralinos and sleptons.  

Since the decay length of the NLSP depends on $\sqrt{F}$ and is effectively 
unconstrained, the NLSP can decay inside or outside of the detector, 
so all possible lifetime topologies are searched for. With increasing decay 
length, the event signatures range from energetic leptons or photons and 
missing energy due to the undetected gravitino, to tracks with large impact 
parameters, kinked tracks, or heavy stable charged particles.

\begin{figure}
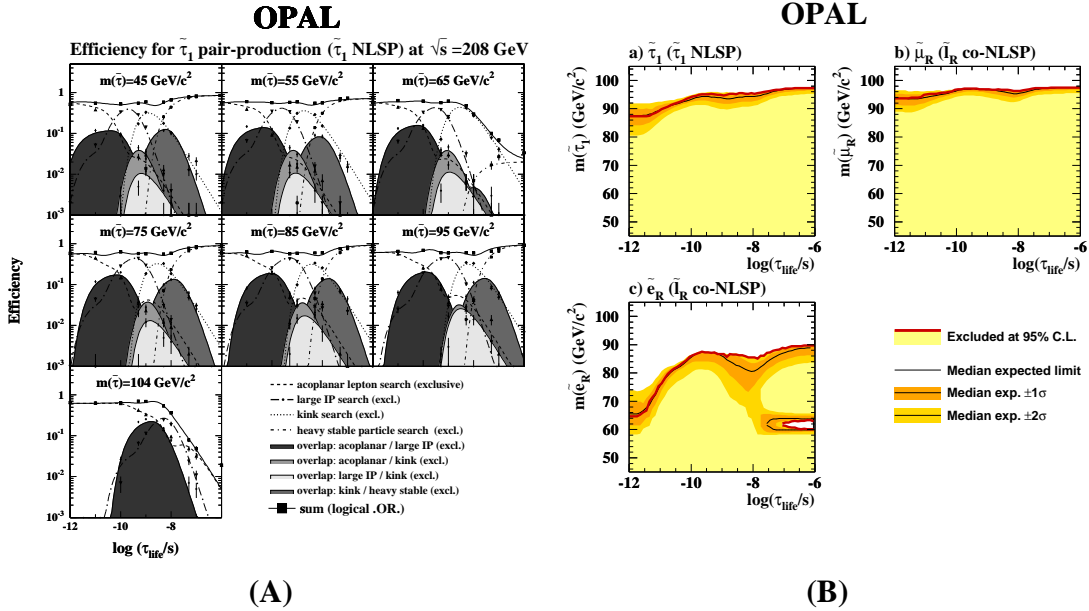

\begin{tabular}{cc}
\includegraphics[scale=0.4]{pr409_06.epsi}&
\includegraphics[scale=0.36]{pr409_19.epsi}\\
\bf{(A)} & \bf{(B)}
\end{tabular}
\caption{{\bf (A)} Efficiencies for stau pair-production at 
$\sqrt{s}=208$~GeV. The symbols represent the efficiencies for ten simulated 
lifetimes while the curves show the interpolating efficiency functions of 
the searches for promptly decaying staus (dashed), large impact parameters 
(long dash-dotted), kinks (dotted) and stable staus (dash-dotted) together 
with the overlap efficiencies (filled areas). The total efficiency is shown 
by full line. {\bf (B)} Observed and expected lower mass limits for 
pair-produced staus in the stau NLSP (a) and smuons (b), selectrons (c) in 
the slepton co-NLSP scenario as a function of the particle lifetime using the 
direct $\slept^+\slept^-$ search. For staus the observed and expected lower 
limit are identical in the stau NLSP scenario and in the slepton co-NLSP 
scenario. The mass limits are valid for a messenger index N$\leq 5$. For the 
stau NLSP and slepton co-NLSP scenarios, the NLSP mass limits are set by the 
stau mass limit and by the smuon mass limit, respectively.}
\label{fig:gmsb}
\end{figure}

In total 14 different selections, each incorporating several signature 
variations, some based on the analyses described in~\cite{opal:old}, 
are implemented to cover all the GMSB topologies: slepton NLSP, neutralino 
NLSP, direct and cascade production, all lifetimes. In order to obtain 
lifetime independent results, the results from the various lifetime 
topologies are combined, with special attention to study the overlaps among 
the many channels. To achieve a good description of the selection 
efficiencies over the whole mass and lifetime range at all center-of-mass 
energies, without generating an excessive number of Monte Carlo samples, an 
interpolating function is determined. On Fig~\ref{fig:gmsb} it is 
demonstrated how the different selections contribute to the signal detection 
efficiency as a function of the NLSP lifetime.

None of the searches shows evidence for SUSY particle production. To interpret
the results, a detailed scan of the minimal GMSB parameter space is performed
with the gravitino mass fixed to 2 eV, corresponding to $\sqrt{F} \approx
100$~TeV, motivated by the requirement that the branching ratio of the
next-to-NLSP to the gravitino is small. If that is the case, the
cross-sections and branching ratios do not depend on the gravitino mass. One
should note that $\sqrt{F}$ can be eliminated from the scan as all limits are
computed independent of the NLSP lifetime, and $\sqrt{F}$ has no significant
effect on other particle masses. 

"Model independent" cross-section limits are derived for each topology as a
function of the NLSP lifetime. For direct NLSP decays, this is done by  
taking the worst limit for a given NLSP mass from the generated GMSB 
parameter scan points. For cascade channels, the cross-section evolution is 
assumed to be $\beta/s$ for spin-1/2 and $\beta^3/s$  for scalar SUSY 
particles, respectively, and the highest bound for all intermediate particle 
masses is retained. The maximum limit valid for all lifetimes is then quoted 
as the "lifetime independent" cross-section limit. In the neutralino NLSP 
scenario this is typically better than 0.04 pb for direct NLSP production, 
0.1 pb for selectron and smuon production, 0.2 pb for stau production and 0.3 
pb for chargino production. In the stau and slepton co-NLSP scenarios, the 
limit on direct NLSP production is 0.05 pb for smuons, 0.1 pb for selectrons 
and staus. For the cascade decays the bounds are typically better than 0.1 pb 
for neutralino, 0.2 for chargino and in the stau NLSP scenario 0.4 for 
selectron and smuon production.

The cross-section limits can be turned into constraints on the NLSP
mass. For sleptons, the lowest mass limits are found for very short
lifetimes, except for selectrons, as shown in Figure~\ref{fig:gmsb}, where 
searches using d$E$/d$x$ measurements lose efficiency for particles with
momenta around 65 GeV. The lifetime independent limits are 
$m_{\tilde\mathrm{e}_\mathrm{R}} > 60.1$~GeV, $m_{\tilde\mu_\mathrm{R}} >
93.7$~GeV and  $m_{\tilde\tau_1} > 87.4$~GeV. The limit on the stau mass is the
same  in the stau and the slepton co-NLSP scenarios. In the slepton co-NLSP
scenario, the best limit can be used to derive a universal limit on the 
slepton masses $m_{\tilde\ell} = m_{\tilde\mu_\mathrm{R}} - m_\tau
> 91.9$~GeV, where by definition the mass differences between the different
slepton flavors are smaller than the lepton masses. For neutralino NLSP, no
lifetime independent NLSP mass limit can be set directly. For short lifetimes
($\tau < 10^{-9}$~s) a mass limit of 96.8 GeV is derived. For the first time 
limits on the production cross-section for all GMSB search topologies, 
including cascade, are presented.

\begin{figure}
\includegraphics[height=.45\textheight]{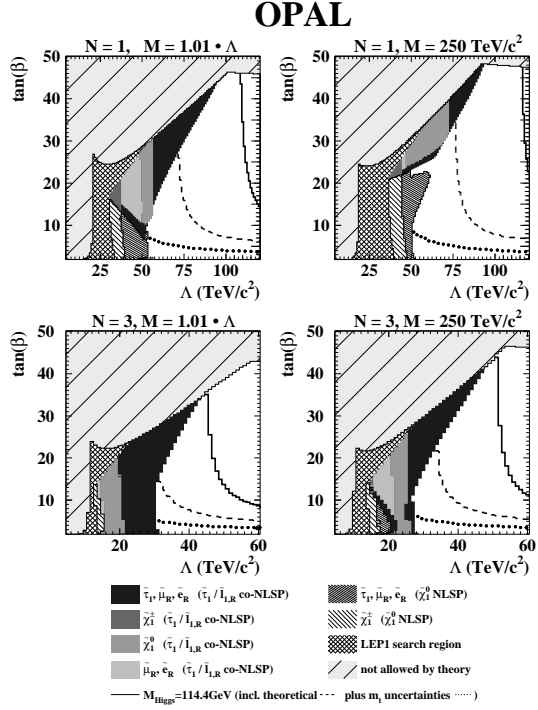}
\caption{Examples of regions in the $\Lambda-\tan{\beta}$ plane excluded by 
pair-production searches for different particles, with sign$(\mu)>0$ and 
valid for any NLSP lifetime for four different sets of parameters, $N=1$ or 
$3$ and $M=1.01\cdot\Lambda$ or $250\,$TeV/$c^2$.}
\label{fig:mGMSB}
\end{figure}

The GMSB parameter space is constrained by our results as shown in
Figure~\ref{fig:mGMSB} for $N=1$, $M=1.01\cdot\Lambda$ and sign($\mu$)>0. The
universal SUSY mass scale is $\Lambda> 40, 27, 21, 17, 15$~TeV for messenger
indices $N=1, 2, 3, 4, 5$, respectively, independent of $M, \tan\beta$,
sign($\mu$) and the NLSP lifetime ($\sqrt{F}$). The constraints on $\Lambda$
imply lower limits on the neutralino mass in the neutralino NLSP scenario:
$m_{\tilde\chi_1^0} > 53.5$~GeV for N=1 and $m_{\tilde\chi_1^0} > 94.0$~GeV 
for N=5, independent of the lifetime.

%
%

\def\Journal#1#2#3#4{{#1} {\bf #2} (#3) #4}


\begin{thebibliography}{9}
\bibitem{susy}
H.P.~Nilles, \Journal{Phys. Rept.}{110}{1984}{1}.
%
\bibitem{theo}
G.F.~Giudice and R.~Rattazzi, \Journal{Phys. Rept.}{322}
{1999}{419},
S.~Ambrosanio, G.D.~Kribs and S.P.~Martin, \Journal{Phys. 
Rev. D}{56}{1997}{1761},
S.~Dimopoulos, S.~Thomas and J.D.~Wells, 
\Journal{Nucl. Phys. B}{488}{1997}{39}.
%
\bibitem{opal:old}
OPAL Collaboration, G.~Abbiendi {\it et al.},
Phys.\ Lett.\ B {\bf 572} (2003) 8,
%
OPAL Collaboration, G.~Abbiendi {\it et al.},
Eur.\ Phys.\ J.\ C {\bf 32} (2004) 453,
%
OPAL Collaboration, G.~Abbiendi {\it et al.},
Phys.\ Lett.\ B {\bf 602} (2004) 167,
%
OPAL Collaboration, G.~Abbiendi {\it et al.}, 
Eur.\ Phys.\ J.\ C {\bf 35} (2004) 1,
%
OPAL Collaboration, G.~Abbiendi {\it et al.}
Eur.\ Phys.\ J.\ C {\bf 33} (2004) 173.
%
\bibitem{opal_gmsb}
  OPAL Collaboration, G.~Abbiendi {\it et al.},
  Eur.\ Phys.\ J.\ C {\bf 46}, 307 (2006).
\end{thebibliography}
\end{document}